\documentclass[article,twocolumn,
superscriptaddress,
amsmath,amssymb,
aps,
pra,
floatfix,
]{revtex4-1}

\usepackage[normalem]{ulem}

\usepackage{graphicx}
\usepackage{xfrac}
\usepackage{amsmath}
\usepackage{mathtools}

\usepackage{epsfig}
\usepackage[colorlinks=true,citecolor=blue,linkcolor=blue,urlcolor=blue]{hyperref}
\usepackage{MnSymbol}
\usepackage{siunitx}
\usepackage{float}

\usepackage{graphicx}
\usepackage{dcolumn}
\usepackage{bm}

\usepackage{tikz,xcolor,hyperref}

\definecolor{lime}{HTML}{A6CE39}
\DeclareRobustCommand{\orcidicon}{
	\begin{tikzpicture}
	\draw[lime, fill=lime] (0,0) 
	circle [radius=0.16] 
	node[white] {{\fontfamily{qag}\selectfont \tiny ID}};
	\draw[white, fill=white] (-0.0625,0.095) 
	circle [radius=0.007];
	\end{tikzpicture}
	\hspace{-2mm}
}
\foreach \x in {A, ..., Z}{\expandafter\xdef\csname orcid\x\endcsname{\noexpand\href{https://orcid.org/\csname orcidauthor\x\endcsname}
			{\noexpand\orcidicon}}
}

\begin{document}

\title{Sensitivity to New Physics of Isotope Shift Studies using the Coronal Lines of Highly Charged Calcium Ions}
	
\author{Nils-Holger Rehbehn\orcidA{}}
\email[]{nils.rehbehn@mpi-hd.mpg.de}
\affiliation{Max-Planck-Institut f\"ur Kernphysik, D--69117 Heidelberg, Germany}

\author{Michael K. Rosner\orcidB{}}
\affiliation{Max-Planck-Institut f\"ur Kernphysik, D--69117 Heidelberg, Germany}

\author{Hendrik Bekker\orcidK{}}
\affiliation{Max-Planck-Institut f\"ur Kernphysik, D--69117 Heidelberg, Germany}
\affiliation{Helmholtz-Institut Mainz, Johannes Gutenberg University, D--55128 Mainz, Germany}

\author{Julian C. Berengut\orcidC{}}
\affiliation{School of Physics, University of New South Wales, Sydney, New South Wales 2052, Australia}
\affiliation{Max-Planck-Institut f\"ur Kernphysik, D--69117 Heidelberg, Germany}

\author{Piet O.~Schmidt\orcidD{}}
\affiliation{Physikalisch--Technische Bundesanstalt, D--38116 Braunschweig, Germany}
\affiliation{Leibniz Universit\"at Hannover, D--30167 Hannover, Germany}

\author{Steven A. King\orcidE{}}
\affiliation{Physikalisch--Technische Bundesanstalt, D--38116 Braunschweig, Germany}

\author{Peter Micke\orcidF{}}
\affiliation{Physikalisch--Technische Bundesanstalt, D--38116 Braunschweig, Germany}
\affiliation{Max-Planck-Institut f\"ur Kernphysik, D--69117 Heidelberg, Germany}

\author{Ming Feng Gu}
\affiliation{Space Science Laboratory, University of California, Berkeley, CA 94720, USA}

\author{Robert M\"uller\orcidH}
\affiliation{Physikalisch--Technische Bundesanstalt, D--38116 Braunschweig, Germany}
\affiliation{Technische Universit\"at Braunschweig, D--38106 Braunschweig, Germany}

\author{Andrey Surzhykov\orcidI{}}
\affiliation{Physikalisch--Technische Bundesanstalt, D--38116 Braunschweig, Germany}
\affiliation{Technische Universit\"at Braunschweig, D--38106 Braunschweig, Germany}
\affiliation{Laboratory for Emerging Nanometrology Braunschweig, D-38106 Braunschweig, Germany}
	
\author{Jos\'e R. {Crespo L\'opez-Urrutia}\orcidJ{}}
\email[]{crespojr@mpi-hd.mpg.de}
\affiliation{Max-Planck-Institut f\"ur Kernphysik, D--69117 Heidelberg, Germany}

\date{\today}
	
\begin{abstract}
Promising searches for new physics beyond the current Standard Model (SM) of particle physics are feasible through isotope-shift spectroscopy, which is sensitive to a hypothetical fifth force between the neutrons of the nucleus and the electrons of the shell. Such an interaction would be mediated by a new particle which could in principle be associated with dark matter. In so-called King plots, the mass-scaled frequency shifts of two optical transitions are plotted against each other for a series of isotopes. Subtle deviations from the expected linearity could reveal such a fifth force. Here, we study experimentally and theoretically six transitions in highly charged ions of Ca, an element with five stable isotopes of zero nuclear spin. Some of the transitions are suitable for upcoming high-precision coherent laser spectroscopy and optical clocks. Our results provide a sufficient number of clock transitions for -- in combination with those of singly charged Ca$^+$ -- application of the generalized King plot method. This will allow future high-precision measurements to remove higher-order SM-related nonlinearities and open a new door to yet more sensitive searches for unknown forces and particles.
\end{abstract}
	
\maketitle
	
Since its inception, the Standard Model (SM) of particle physics has been the cornerstone of our understanding of nature, but since it leaves open fundamental questions about dark matter, dark energy, mass hierarchy, and others, it is considered still incomplete, and therefore new physics (NP) is sought after.

By exploiting the unrivalled accuracy of laser spectroscopy, atomic physics offers unique sensitivity for searches for physics beyond the SM (for a review see \cite{Safronova2018}). Feeble non-gravitational interactions between normal matter and, e.~g., dark matter, would result in changes of atomic and molecular energy levels. Depending on the dark matter candidate and the properties of the field representing it, oscillations \cite{arvanitaki_searching_2015, stadnik_searching_2015}, drifts \cite{stadnik_can_2015} or transient changes \cite{derevianko_hunting_2014} in energy levels can occur. Optical atomic clocks \cite{ludlow_optical_2015} with up to 18 digits of accuracy have already been employed to put bounds on the mass of dark matter candidates \cite{Kennedy_2020, wcislo_experimental_2016, wcislo_new_2018, roberts_search_2020}.

Recently, a complementary approach \cite{Delaunay2017, Berengut2018} proposes probing the existence of a hypothetical fifth force coupling electrons and neutrons by means of isotope-shift (IS) spectroscopy \cite{Solaro2020, counts_observation_2020}. Isotopic perturbations of optical transitions are dominated by two effects: (i) the nuclear recoil (mass shift, MS), and (ii) the modification of the electron-nucleus interaction potential by the nuclear charge distribution (field shift, FS). The dependence on the mostly poorly known nuclear charge distribution is eliminated by measuring \cite{Knollmann2019, Manovitz2019, Miyake2019, gebert_precision_2015-1, shi_unexpectedly_2017, muller_collinear_2020} two different transitions and using a so-called King-plot \cite{king_1963}. This yields a linear relationship in first order between the two transition frequencies. 
A fifth force coupling electrons and neutrons would break this linearity \cite{Delaunay2017,Frugiuele2017,Flambaum2018,Fichet2018,Berengut2018}. Caution is, however, needed, since nonlinearity can also arise from higher-order SM effects \cite{Flambaum2018,Berengut2018}, which then cloud the NP effects. 
To separate them, high precision atomic and nuclear structure calculations for the former are required \cite{Yerokhin2020,Reinhard2020,tanaka_relativistic_2020,Mikami2017,Flambaum2018}, which are feasible for few-electron systems such as highly charged ions (HCI). Alternatively, a generalized King plot (GKP) proposed by \citet{Mikami2017} and further developed by  \citet{berengut_2020} employs measurements of additional electronic transitions for eliminating the impact of higher-order SM effects on such NP searches.

\begin{figure*} 
\centering
\includegraphics[width=\linewidth]{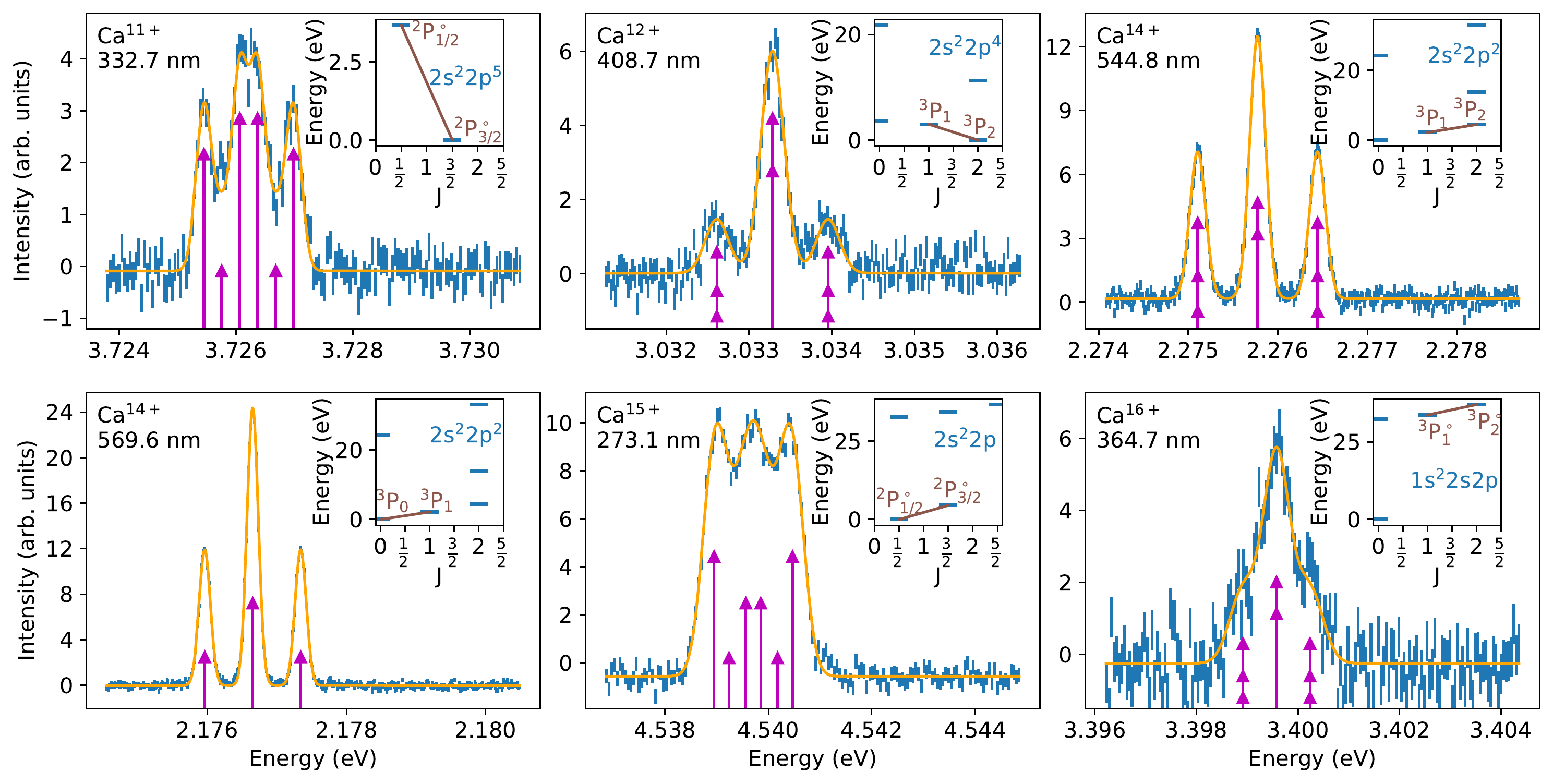}
\caption{Fits to the Zeeman components of the studied transitions. Magenta arrows mark their positions and relative intensities. The inset level diagrams with total angular momentum $J$ were calculated with \textsc{fac}~\cite{gu_2008}.}
\label{fig:zeeman_fits}
\end{figure*}

\begin{table*}
\centering
\caption{Forbidden optical fine-structure transitions in highly charged Ca ions: Measured energies, vacuum wavelengths have been observed, transition probabilities $A_{ki}$ and theoretical energy values were calculated in this work with \textsc{fac}~\cite{gu_2008} and \textsc{ratip/grasp}~\cite{fritzsche_2012}. Their respective SM electronic sensitivity coefficients K and F have been calculated with \textsc{fac}. For comparison, the $729\,\mathrm{nm}$ Ca$^+$ transition has been included with values from the NIST database \cite{NIST_database}.}
\begin{tabular}{l|llll|l|l|l|l|l|l|l|l}
& \multicolumn{4}{c|}{}&	\multicolumn{2}{c|}{Observed} & \multicolumn{2}{c|}{\textsc{fac}} & \multicolumn{2}{c|}{\textsc{ratip/grasp}} & \multicolumn{2}{c}{Elec. coeff.} \\
Ion & \multicolumn{4}{c|}{Transition} &	Energy (eV) & Wavel. (nm) & En. (eV) & $A_{ki}$ (s$^{-1}$)&En. (eV) & $A_{ki}$ (s$^{-1}$) & K (eV u) & F (eV/fm$^{2}$)  \\
\hline
Ca$^{11+}$&$2s^22p^5$&$ ^2\!P^\circ_{1/2} $&-&$ ^2\!\!P^\circ_{3/2}$ & 3.7262192(28) & 332.73458(25) & 3.716(3) & 483(1) & 3.718(1) & 483.9(1) & -$2.85 \times 10^{-3}$ & -$104 \times 10^{-9}$ \\
Ca$^{12+}$&$2s^22p^4$&$ ^3\!P_1 $&-&$ ^3\!\!P_2$ & 3.0332843(19) & 408.74572(26) & 3.022(4) & 316(1) & 3.012(2) & 312(1) & -$2.43 \times 10^{-3}$ & -$123 \times 10^{-9}$  \\ 
Ca$^{14+}$&$2s^22p^2$&$ ^3\!P_2 $&-&$ ^3\!\!P_1$  & 2.2757757(7) & 544.79971(16) & 2.31(2) & 83(2) & 2.32(2) & 83(2) & -$1.90 \times 10^{-3}$ & -$196 \times 10^{-9}$  \\ 
Ca$^{14+}$&$2s^22p^2$&$ ^3\!P_1 $&-&$ ^3\!\!P_0$ & 2.1766536(10) & 569.60923(26) & 2.15(5) & 91(6) & 2.06(3) & 81(3) & -$1.83 \times 10^{-3}$ & -$21 \times 10^{-9}$  \\ 
Ca$^{15+}$&$2s^22p$&$ ^2\!P^\circ_{3/2} $&-&$ ^2\!\!P^\circ_{1/2}$ & 4.5397089(27) & 273.11046(16) & 4.60(6) & 459(17) & 4.5352(2) & 439.0(1) & -$4.96 \times 10^{-3}$ & -$577 \times 10^{-9}$  \\ 
Ca$^{16+}$&$1s^22s2p$&$ ^3\!P^\circ_2 $&-&$ ^3\!\!P^\circ_1$& 3.3995766(74) & 364.70482(79) & 3.3839(8) & 272.2(2) & 3.392(3) & 274(1) & -$2.76 \times 10^{-3}$ & -$134 \times 10^{-9}$ \\ 
Ca$^{+}$&$3p^63d$-$4s$&$ ^2\!D_{5/2} $&-&$ ^2\!\!S_{1/2}$& 1.699932 \cite{NIST_database} & 729.348 \cite{NIST_database} & 1.88(6) & 1.7(2) & 2.0228(4) & - & -$3.13 \times 10^{-3}$ & -$1.31 \times 10^{-6}$
\end{tabular}
\label{tab:results}
\end{table*}

Recently, optical-clock-like spectroscopy of HCI was demonstrated by means of sympathetic laser cooling and quantum logic operations in a linear Paul trap \cite{micke_2020}. Relative fractional uncertainties as low as $10^{-16}$ can be achieved through absolute frequency measurements, limited by the SI-second. For optical transition frequencies of the order of $500\,\mathrm{THz}$ ($2\,\mathrm{eV}$) this corresponds to an absolute uncertainty of the order of $100\,\mathrm{mHz}$ ($0.4\,\mathrm{feV}$), which is roughly a thousand times smaller than the natural linewidths of magnetic-dipole transitions in HCI. Nevertheless, this level of subdivision is possible using techniques employed by atomic clocks.
With such precision, it would be possible to constrain the NP \cite{berengut_2020} beyond the limits set in neutron \cite{barbieri_1975,leeb_1992,nesvizhevsky_2008,pokotilovski_2006} and electron scattering \cite{adler_1974}, and fifth-force studies \cite{bordag_2001,bordag_2009}. 

In this Letter, we experimentally determine accurate wavelengths for six magnetic dipole (M1) forbidden lines of Ca ions in the charge stages $11+$ through $16+$, and identify which ones transition directly to the electronic ground state and thus are suitable for coherent laser spectroscopy \cite{schmidt_2005} IS measurements. We perform calculations of their IS and use them to construct King plots (KP) as well as generalized King plots (GKP) to suppress higher-order SM terms, to see their effects and find the most suitable combination of transitions for the search of NP influences.

Calcium ions are particularly suitable for King-plot studies since this element has five stable isotopes with zero nuclear spin. Additionally, Ca has narrow transitions in various charge states which can be accurately measured by high-precision laser spectroscopy using common optical frequency metrology methods. In the past, the isotope shifts of broad dipole allowed  \cite{Noertershaeuser1998,Mortensen2004,Gorges2015,Hashimoto2011,Shi2016,Gebert2015} as well as narrow transitions \cite{Benhelm2007,Knollmann2019,Solaro2020} in singly charged Ca have been investigated. 
In HCI, the strongly bound outer electrons can undergo fine-structure \cite{Morgan1995,Draganic2003,SoriaOrts2006,Crespo2008,lopez-urrutia_emission_2014,Murata2017,Windberger_2016,Bekker2018} and hyperfine-structure \cite{klaft_precision_1994,Crespo1996PRL,Seelig1998PRL,Crespo1998PRA,Beiersdorfer2001PRA} transitions in the optical range. Some of them contribute to the corona spectrum and are therefore known as \textit{coronal} lines. 
Optical transitions in HCI can also arise from level crossings \cite{Berengut2010,Berengut2012,Windberger2015,Bekker2019}. These involve orbitals of very different character, and thus promise a high sensitivity to NP (see, e.g.~\cite{ong2014optical,berengut_electron-hole_2011,derevianko_highly_2012,Dzuba2015,SafronovaPRA2014,Porsev2020,Cheung2020,Kimura2020}), making them strong candidates for such studies \cite{KozlovRMP2018}. 

To produce Ca HCI, we used the Heidelberg electron beam ion trap (EBIT) \cite{levine_electron_1988,levine_use_1989,crespo_1999}. There, an electron beam emitted by a thermionic cathode is accelerated to energies between $600\,\mathrm{eV}$ and $1200\,\mathrm{eV}$ and strongly compressed by a coaxial 8\,T magnetic field. Depending on its energy, electron impact ionization brings atoms crossing the beam to the desired charge state. Resulting HCI are radially trapped by the negative space charge potential generated by the electron beam and axially by biased drift tubes. 
A small oven is used to evaporate a substance containing Ca. The resulting molecular beam crosses the electron beam and is dissociated there, thus releasing millions of Ca atoms for ionization. Collisions with the electron beam heat up these HCI in the deep trapping potential of the EBIT to temperatures around $10^5$ to $10^6$\,K. These values depend on the charge as well as current density of the beam, and on the axial trapping potential. Less current or a shallow trap lead to lower temperature and smaller Doppler width, however at the cost of a weaker signal. The trap is periodically dumped every few seconds in order to avoid a slow accumulation of undesired ions of barium and tungsten, elements which are constituents of the thermoionic cathode and slowly evaporate from there.

In the trap, the forbidden optical transitions of interest are excited by electron impact. Using four lenses and several mirrors, we project a rotated image of the horizontal ion cloud onto the vertical entrance slit of a 2-meter Czerny-Turner spectrometer \cite{Bekker2018,Bekker2019} equipped with a cryogenically cooled CCD camera. We calibrate the spectral dispersion function through a polynomial fit of the line positions of known transitions \cite{NIST_database}, emitted by suitable hollow cathode lamps. Spectra of such lamps are automatically taken before and after each of the approximately 30-minute-long exposures needed for the HCI lines. The two-dimensional spectral images are cleaned from cosmic events, corrected from optical aberrations and calibrated, as explained in detail in Refs. \cite{Windberger2015,Bekker2018,Bekker2019}. Transition wavelengths are determined by fitting Gaussians to the resolved Zeeman components, taking calculated Clebsch-Gordan coefficients  as initial parameters for the $g$-factors. Results are shown in figure~\ref{fig:zeeman_fits} with the respective level diagrams of the transitions, and summarized in Table~\ref{tab:results}. 
We estimate the uncertainty as the square root of the quadratic summation of the uncertainties of both the Zeeman line fit (for the Ca$^{16+}$ transition the largest contribution) and the dispersion-function calibration, which dominates for the two Ca$^{14+}$ transitions.
The four transitions of the Ca$^{11+}$, Ca$^{12+}$ and Ca$^{14+}$ were observed in the solar corona with roughly hundredfold larger uncertainties \cite{Jefferies_1971}.

For the observed $p_{j1}\rightarrow p_{j2}$ transitions in the charge states from $\mathrm{Ca}^{11+}$ to $\mathrm{Ca}^{16+}$, as well as the $\mathrm{Ca}^{+}$ $s_{j1}\rightarrow d_{j2}$ transition, we calculated theoretical transition energies and their dependence on hypothetical NP with the Flexible Atomic Code (\textsc{fac}) version 1.1.5 \cite{gu_2008} and crosscheck with \textsc{ratip/grasp} \cite{fritzsche_2012}. For this, a fictitious mediator $\Phi$ is modeled by a Yukawa central potential $V_\mathrm{\Phi}(r)$ and introduced as a perturbation to the electromagnetic field acting upon the electrons:
\begin{equation}
V_\mathrm{\Phi}(r)= y_\mathrm{e} y_\mathrm{n} (A-Z) \frac{\hbar \mathrm{c}}{4 \pi r} \exp \left( -\frac{c}{\hbar} \cdot m_\Phi \cdot r\right).
\label{eq:yukawa}
\end{equation}
Here, $\hbar$ is the reduced Planck constant and $c$ the speed of light. The coupling strength is defined as $y_\mathrm{e} y_\mathrm{n}$, where $y_\mathrm{e}$ and $y_\mathrm{n}$ are the couplings of $\Phi$ to electrons and neutrons, respectively. $A$ is the mass number and $Z$ is the nuclear charge. For the present estimates, we set a value of $y_\mathrm{e}y_\mathrm{n}=1\times 10^{-13}$, which is below the limits of current exclusion plots, such as shown in Ref.~\cite{berengut_2020}. 
The Yukawa range is set by the mediator mass $m_\Phi$: a lighter particle has a longer-range effect than a heavier one. 

\begin{figure}
\centering
\includegraphics[width=\linewidth]{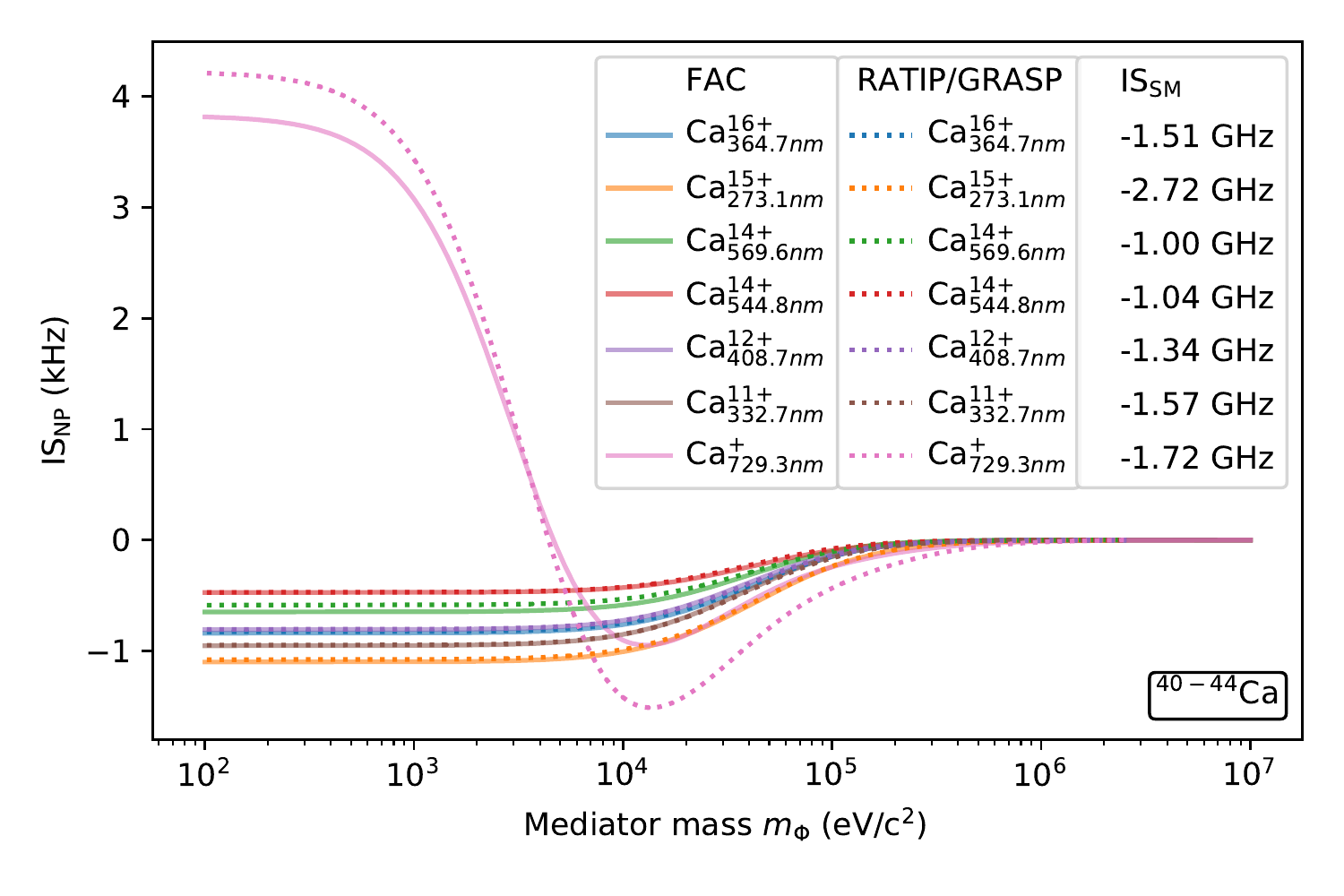}
\caption{Effect of the NP particle on the isotope shift (IS$_\mathrm{NP}$) predicted using \textsc{fac} and \textsc{ratip/grasp} for the Ca isotope pair (40,44) as a function of the mediator mass for a coupling strength $y_\mathrm{e}y_\mathrm{n}=1\times 10^{-13}$. The dominant SM contribution (IS$_\mathrm{SM}$) was subtracted from the total IS to make the smaller IS$_\mathrm{NP}$ contributions visible.}
\label{fig:FAC_predictions}
\end{figure}

\begin{table}
\centering
\caption{NP electronic coefficients X($m_\Phi$) in eV calculated with \textsc{fac} and \textsc{ratip/grasp} for different mediator masses $m_\Phi$.}
\begin{tabular}{l|l|l|l|l|l|l|l|l}
&\multicolumn{8}{c}{$m_\Phi$}\\
&\multicolumn{2}{c|}{10$^3$ ${\mathrm{eV}}/{\mathrm{c}^2}$} & \multicolumn{2}{c|}{10$^4$ ${\mathrm{eV}}/{\mathrm{c}^2}$} & \multicolumn{2}{c|}{10$^5$ ${\mathrm{eV}}/{\mathrm{c}^2}$} & \multicolumn{2}{c}{10$^6$ ${\mathrm{eV}}/{\mathrm{c}^2}$} \\
\hline
Trans. & \textsc{fac} & \textsc{grasp} & \textsc{fac} & \textsc{grasp} & \textsc{fac} & \textsc{grasp} & \textsc{fac} & \textsc{grasp}\\
\hline 
Ca$^{11+}$ & 9.85 & 9.81 & 8.82 & 8.80 & 1.54 & 1.67 & 0.0163 & 0.0252\\
Ca$^{12+}$ & 8.36 & 8.33 & 7.51 & 7.51 & 1.33 & 1.45 & 0.0139 & 0.0221\\
Ca$_{544}^{14+}$ & 4.89 & 4.88 & 4.45 & 4.40 & 0.96 & 0.80 & 0.0203 & 0.0087\\
Ca$_{569}^{14+}$ & 6.71 & 6.06 & 6.07 & 5.51 & 1.07 & 1.07 & 0.0065 & 0.0126\\
Ca$^{15+}$ & 11.4 & 11.2 & 10.4 & 10.2 & 2.47 & 2.44 & 0.0605 & 0.0573\\
Ca$^{16+}$ & 8.69 & 8.56 & 7.90 & 7.78 & 1.52 & 1.51 & 0.0177 & 0.0177\\
Ca$^{+}$   & -31.9 & -35.3 & 9.37 & 14.8 & 2.51 & 4.52 & 0.119 & 0.223
\end{tabular}
\label{tab:Xis}
\end{table}
	
In first-order perturbation theory, the IS is defined for transition $i$ as the sum of the SM mass and field shift, as well as the shift induced by the NP mediator:
\begin{equation}
\mathrm{IS}=\mathrm{IS}_\mathrm{SM}+\mathrm{IS}_\mathrm{NP}\ \equiv \ \delta\nu_i^a=K_i \mu_{a} + F_i \delta\langle r_{a}^2 \rangle + y_\mathrm{e} y_\mathrm{n} X_i \gamma_{a},
\label{eq:IS}
\end{equation}
where $a$ is the isotope pair $(A, A_r)$, thus $\delta\nu_i^a=\nu_i^{A}-\nu_i^{A_r}$ is the difference of transition $i$ between isotope $A$ and reference isotope $A_r$; $K_i$, $F_i$ and $X_i$ are electronic constants of the shift terms, $\mu_a=1/m_{A} - 1/m_{A_r}$ is the difference of the inverse isotope masses and $\delta\langle r_a^2\rangle=\langle r_{A}^2\rangle-\langle r_{A_r}^2\rangle$ is the difference of the mean square nuclear charge radii. The NP part depends on the isotopes with $\gamma_a=(A-Z)-(A_r-Z)=A-A_r$ and $y_\mathrm{e} y_\mathrm{n} \propto V_\mathrm{\Phi}$ (see equation~\ref{eq:yukawa}) defining the coupling strength with regards to the coupling range.

Figure~\ref{fig:FAC_predictions} shows the IS of the studied transitions induced by NP for the Ca isotope pair (40,44); the associated electronic coefficients are listed in table~\ref{tab:Xis}. To highlight the effects of the NP particle, the dominating SM contribution was subtracted from the total IS. 
For comparison, we also included the Ca$^{+}$ $s_{j1}\rightarrow d_{j2}$ transition. It shows a larger shift than the $p_{j1}\rightarrow p_{j2}$ transitions due to the difference in $l$ between the initial and final state. 
One advantage of HCI is that their reduced number of electrons makes calculations more easily converge than those for neutrals or singly charged ions, as apparent from the differences between FAC and RATIP/GRASP for Ca$^+$.
	
\begin{figure}
	\centering
	\includegraphics[width=\linewidth]{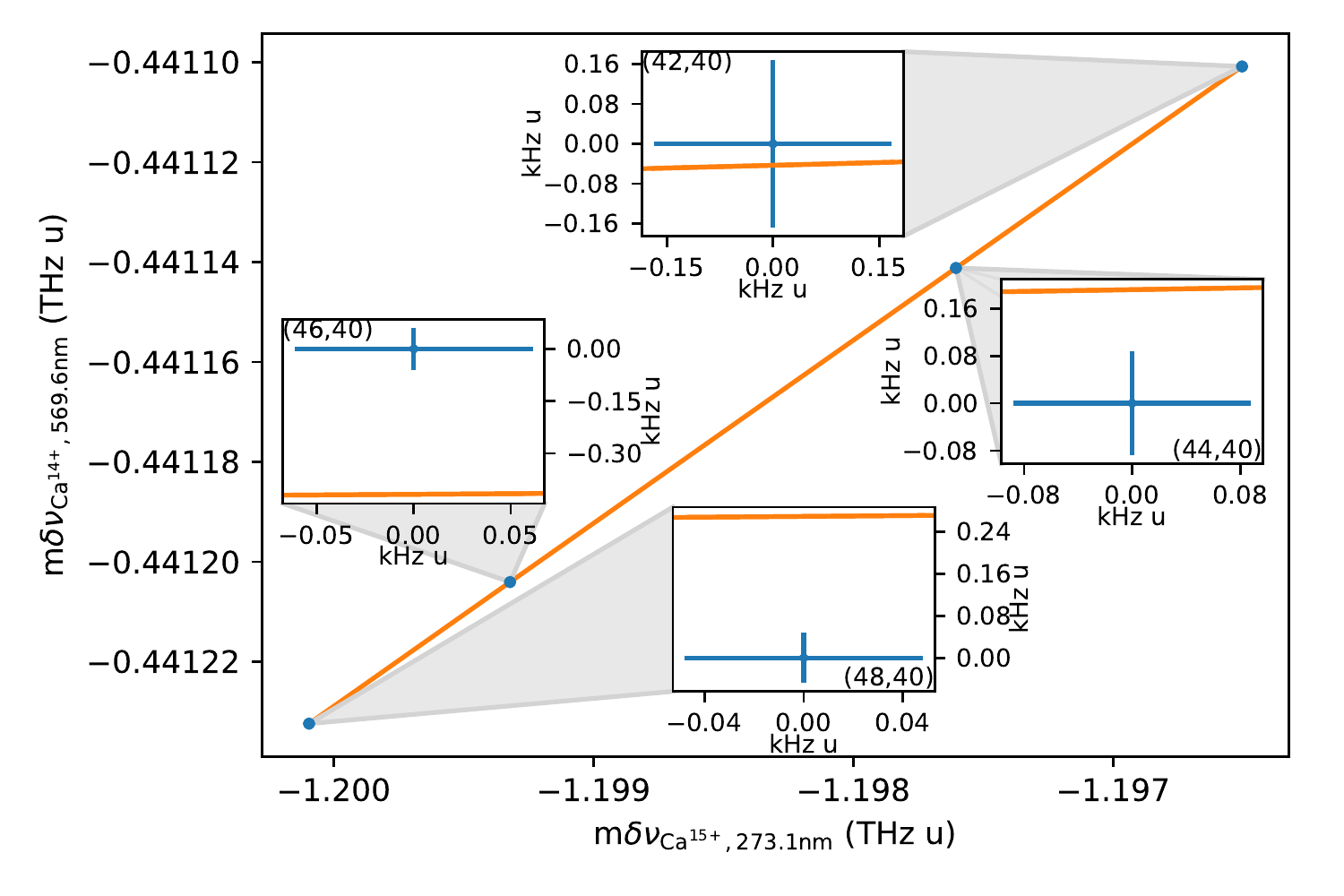}
	\caption{King's plot of \textsc{fac}-calculated transitions in Ca$^{15+}_{273.1\mathrm{nm}}$ and Ca$^{14+}_{569.6\mathrm{nm}}$. It assumes a mediator with mass $m_\Phi=10^5\,\mathrm{eV}/\mathrm{c}^2$ and a coupling strength $y_\mathrm{e}y_\mathrm{n}=1\times 10^{-13}$. The error bars depict a $100\,\mathrm{mHz}$ measurement uncertainty, the mass uncertainty is neglected.
	}
	\label{fig:FAC_KP}
\end{figure}

To analyze the IS without an accurate knowledge of the change of the nuclear charge radius $\langle r^2\rangle$, one can use the approach of King \cite{king_1963}, where the isotope shifts of two transitions in different isotope pairs $a$, divided by the mass parameter $\mu_a$ (denoted as $m$ for modified), are plotted against each other. A linear behavior along all points is expected from first-order perturbation theory, while nonlinearities would arise from either higher order effects from the SM, or from NP parts:
\begin{equation}
\begin{aligned}
    \mathrm{m}\delta\nu_2^a=& \frac{F_2}{F_1}\cdot\mathrm{m}\delta\nu_1^a+\left(K_2-\frac{F_2}{F_1}K_1\right)\\ &+ y_\mathrm{e} y_\mathrm{n}\cdot\left(X_2  - \frac{F_2}{F_1} X_1 \right)\cdot \mathrm{m}\gamma_a .
    \label{eq:KP_linearity}
\end{aligned}
\end{equation}
Here, the first two terms represent the linear behavior between the isotope shifts of the two transitions $i=2$ and $i=1$, from equation~\ref{eq:IS}, for different isotope pairs $a$. The third term, with the variable $\mathrm{m}\gamma_a=\frac{\gamma_a}{\mu_a}$, induces a nonlinearity when $X_i/F_i$ varies for the transitions and mediator masses.
The King plot can be constructed from the \textsc{fac} results, where the SM and NP electronic coefficients can be taken from table~\ref{tab:results} and table~\ref{tab:Xis}, respectively. This is shown for the mediator mass of $m_\Phi=10^5\,\mathrm{eV}/\mathrm{c}^2$ and a coupling strength of $y_\mathrm{e}y_\mathrm{n}=1\times 10^{-13}$ in figure~\ref{fig:FAC_KP}.

\begin{figure}
\centering
\includegraphics[width=\linewidth]{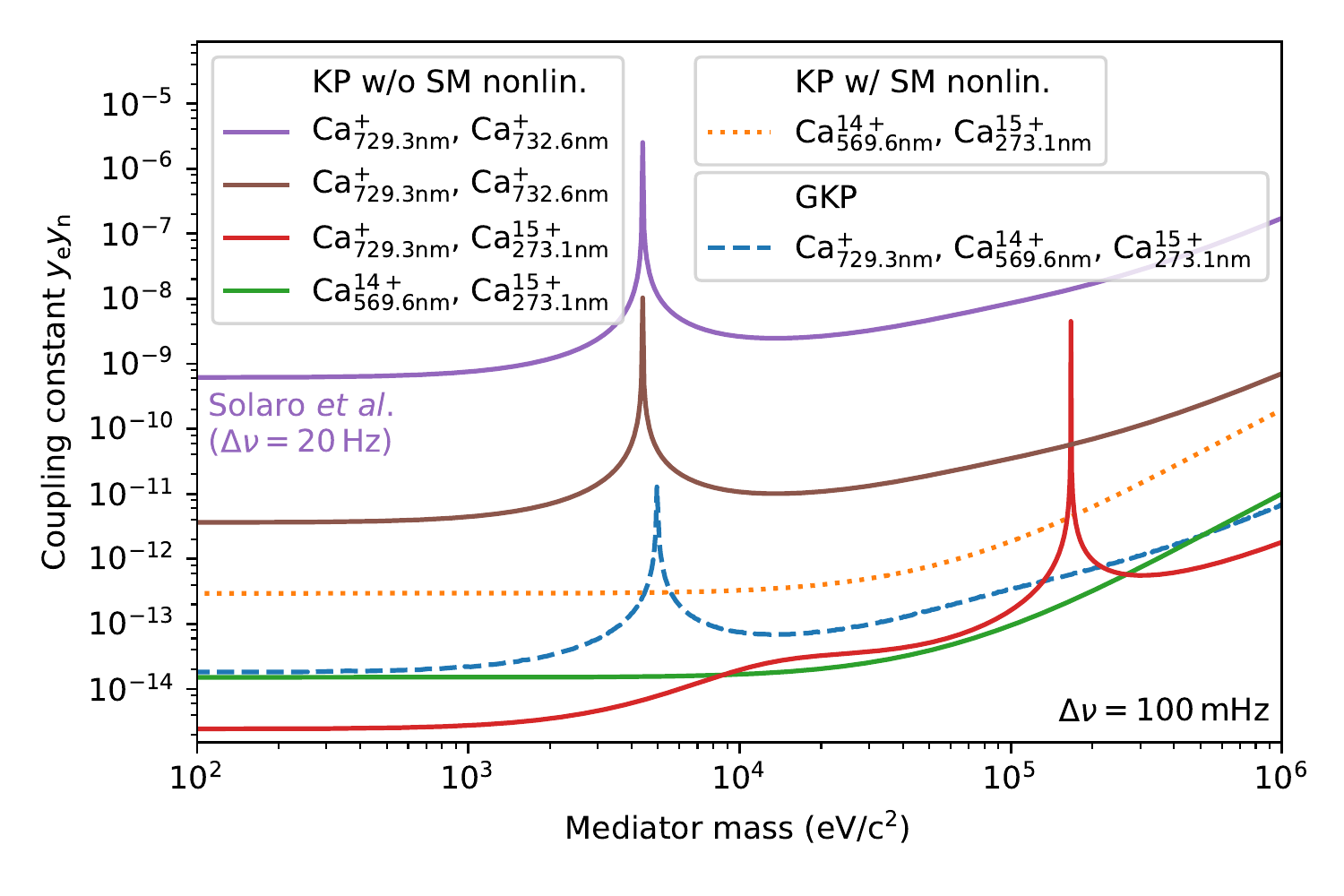}
\caption{Projected bounds on Yukawa interactions excluded by different King plots with measurement errors of $\Delta \nu = 100 \mathrm{mHz}$ and assumed negligible mass uncertainty. Nonlinearities from higher order SM terms limit the bounds which can be placed on the coupling strength. Most of the sensitivity can be recovered by utilizing the generalized King plot (GKP) \cite{berengut_2020}. Calculated with data obtained using \textsc{ratip/grasp}~\cite{fritzsche_2012}.}
\label{fig:exclusion_plot}
\end{figure}

To quantify the chances of detecting NP, we can display the nonlinearity over its error, using the equations from \citet{berengut_2020}. This allows us to plot the lowest possible coupling strength where the nonlinearity can still be resolved with a given measurement uncertainty. Figure~\ref{fig:exclusion_plot} plots this measurement boundary for different mediator masses with an expected measurement uncertainty of $100\,\mathrm{mHz}$ for future coherent laser spectroscopy measurements. The $\mathrm{Ca}^+_{732.6\mathrm{nm}},\mathrm{Ca}^+_{729.3\mathrm{nm}}$ (violet) curve corresponds to the transitions used in \citet{Solaro2020} with a measurement uncertainty of 20~Hz. The brown curve corresponds to this transition pair with the uncertainty level from the present paper. 
As their electronic coefficients are very similar, they are not an ideal pair for the King plot analysis.
Two out of the four transitions ground-state transitions studied here are shown in a King plot. The most promising candidate is the $\mathrm{Ca}^{+}_{729.3\mathrm{nm}},\mathrm{Ca}^{15+}_{273.1\mathrm{nm}}$ pair, but the HCI-only pair $\mathrm{Ca}^{14+}_{569.6\mathrm{nm}},\mathrm{Ca}^{15+}_{273.1\mathrm{nm}}$ shows also a greater chance of finding NP than the $\mathrm{Ca}^+$-only pair. To show potential higher order SM effects, a quadratic mass shift in order of $3\,\mathrm{Hz}$ \cite{Flambaum2018} is added to the HCI pair $\mathrm{Ca}^{15+}_{569.6\mathrm{nm}},\mathrm{Ca}^{14+}_{273.1\mathrm{nm}}$. This term causes additional SM nonlinearities, which greatly limit the bounds that can be placed in the coupling strength (orange). However, it is expected that ongoing calculations of the quadratic mass shift terms in Ca$^+$ and the different Ca HCI will reduce their contributions to the general uncertainty. Furthermore, the problem can be bypassed altogether by using the generalized King plot (blue), as the additional transition is used to separate the higher-order SM effects from the NP effects, similarly to what has been done to obtain equation~\ref{eq:KP_linearity}. It is important to mention that specific knowledge of the higher orders of the SM contributions is not needed here \cite{Mikami2017,berengut_2020}.  

With the higher order SM limitations overcome, we now discuss the required accuracy. A statistical uncertainty of approximately 2~Hz was achieved for a $17\,\mathrm{Hz}$ broad magnetic-dipole (M1) transition in Ar$^{13+}$ \cite{micke_2020} through laser scans of the line. By actively stabilizing that probe laser to the M1 clock transition \cite{Peik_2005,Riis_2004}, frequency data could be continuously acquired for hours or days and thereby an even lower uncertainty reached. This requires cooling the HCI down to ($T\ll~1$~K) for suppression of Doppler broadening and other systematic shifts. Most HCIs lacking laser-cooling transitions, we employ instead sympathetic cooling with a single $^{9}$Be$^{+}$ located in the same trapping potential as the HCI \cite{micke_2020}. 
However, because simultaneously cooling and manipulating two different HCIs with one Be ion in a three-species Coulomb crystal would be very complex, IS frequency measurements are most likely to be performed sequentially. If the primary frequency standard, the Cs fountain, is used as reference, it would limit the fractional accuracy to the low $10^{-16}$ level. This underlies our assumed 100~mHz uncertainty for an individual transition, for which several weeks of continuous averaging would be required (see, for example, \cite{Lange_2021}). A better option would be direct optical-optical comparisons \cite{Rosenband2008,Godun2014,Beloy2020,Doerscher2020,Lange_2021} against an optical frequency standard that is more stable than the Cs standard, yielding frequency differences for the isotopes independent of it. The statistical uncertainty would then be dominated by the comparatively broad natural linewidth of the HCI clock transition. With the system presented in Ref. \cite{king_2021} and a 7~ms-long clock transition interrogation pulse (leading to an interaction-time limited linewidth of 114~Hz), quantum projection noise would cause a statistical uncertainty of approximately $10\,\mathrm{Hz}/\sqrt{\tau}$, where $\tau$ is the averaging time in seconds. The assumed uncertainty of 100~mHz ($0.4\,\mathrm{feV}$) would then require only a few hours of averaging, instead of weeks. If the systematic shifts are controlled at the level which is now standard in optical atomic clocks, their contribution to the overall uncertainty will become negligible.

In future, if the difficulties of using three-ion Coulomb crystals can be overcome, IS could be determined in a single measurement using common or separate spectroscopy lasers. If ions of two or more different isotopes are entangled to prepare decoherence-free, noise-insensitive Hilbert subspaces \cite{Pruttivarasin2015}, the accuracy of the IS frequency comparisons would be further increased, as shown in Ref. \cite{Manovitz2019}.

Our measurements and calculations established six forbidden, laser-accessible transitions in highly charged calcium ions, four of which have a transition to the ground level, and thus are accessible for coherent laser spectroscopy, i.~e., quantum logic spectroscopy. This expands the possibilities for King plot-based IS searches for new physics. 
Taken together with the two already well-studied optical transitions in Ca$^+$ ions, the number of combinations suitable for NP searches grows \cite{Delaunay2017,Berengut2011,Frugiuele2017,Fichet2018,Berengut2018,Flambaum2018} and opens up the use of the generalized King-plot analysis \cite{berengut_2020} for this type of experiment. In this way, limitations by higher-order SM can be overcome, and the NP parameter space can be probed more stringently than currently possible.

\section*{Acknowledgements}
Financial support was provided by the Max-Planck-Gesellschaft and the Physikalisch-Technische Bundesanstalt. We acknowledge support from the Max Planck-Riken-PTB Center for Time, Constants and Fundamental Symmetries, the Deutsche Forschungsgemeinschaft through SCHM2678/5-1 and SU 658/4-1, the collaborative research centers \textquotedblleft SFB 1225 (ISOQUANT)\textquotedblright and \textquotedblleft SFB 1227 (DQ-mat)\textquotedblright, and Germany's Excellence Strategy-EXC-2123/1 QuantumFrontiers-390837967. This project also received funding from the European Metrology Programme for Innovation and Research (EMPIR) cofinanced by the Participating 5 States and from the European Union's Horizon 2020 research and innovation programme (Project No. 17FUN07 CC4C). JCB was supported in this work by the Alexander von Humboldt Foundation and the Australian Research Council (DP190100974).

\bibliography{literature.bib}
	
\end{document}